\documentclass{elsart}
\usepackage{latexsym}
\usepackage{amsmath}
\usepackage{graphicx}
\begin{document}
\begin{frontmatter}
\title{How to distinguish between stick-slip and pure slip oscillations}
\author{J.Szkutnik},
\author{K.Ku{\l}akowski$^*$}


\address{Faculty of Physics and Applied Computer Science, 
AGH University of Science and Technology, al. Mickiewicza 30, 30-059 Cracow, Poland\\
$^*$ Corresponding author. E-mail: kulakowski@novell.ftj.agh.edu.pl}

\begin{abstract} Numerical simulations are performed for the stick-slip motion
in the Burridge-Knopoff model of one block. Calculated amplitude increases with
the driving velocity. We argue that this effect can be a criterion to distinguish 
between the stick-slip and pure slip oscillations.


\end{abstract}
\end{frontmatter}

\section{Introduction}

Understanding and control of frictional properties of matter remains a challenge for a 
wide spectrum of sciences \cite{sbem}. In particular, physical effects which appear in 
the low velocity regime are of interest \cite{hes,braun,alea,kry}. While details of 
rubbing surfaces depend on mechanical properties of the probing system \cite{roz}, 
our understanding of these detail is to be built up from simple models. 

Here we concentrate on the simplest version of the Burridge-Knopoff model \cite{bk,cl} 
of one block, connected with a spring to a driving mechanism moving with a constant 
velocity $v_0$. The block moves on a substrate under forces of spring and friction. 
The core of the problem is the friction force $F$ dependence of the velocity $v$. 
As it was recognized by multiple authors, the origin of the instability of the uniform 
motion is that $F(v)$ decreases at least near $v=0$. This remains true for a chain of 
any number of blocks \cite{vieira}. 

We note that a more detailed model of friction is known \cite{rr,hes} includes the d
ependence of the friction force on the age of contact. This approach, known as Rice-Ruina 
model from names of its authors, has some shortcomings; in its simplest form it does not 
reproduce stable orbits. An argument, somewhat lengthy, was presented by us in 
Ref. \cite{zak}. However, more strong theorem can be proved very briefly, basing on 
the Dulac criterion \cite{glen}. The proof is shown in the Appendix. Here we are 
interested in the stable orbits, then we work with the Burridge-Knopoff model.

It is often assumed that the only alternative for the uniform motion is the stick-slip 
motion, where the block periodically switches from motion to rest and motion again. 
However, as it was recently demonstrated within an analytical approximation, within some 
range of the driving velocity the stick phase disappears, but still the uniform motion 
is not stable \cite{thf}. The motion was termed 'pure-slip oscillations'. The stick-slip, 
pure slip and uniform motion are then three distinct possibilities, one succeeding another 
when the velocity $v_0$ increases. From the results of Ref. \cite{thf}, a criterion can 
be derived to distinguish stick-slip and pure slip phases. The amplitude of the 
oscillations increases with the driving velocity $v_0$ for the stick-slip phase and it 
decreases for the pure slip phase. 

In a recent report \cite{zv}, oscillations of motion of a PMMA polymer sample scratched 
by a conical diamond intender were measured as dependent on the driving velocity and normal 
load. The amplitude and period of oscillations are found to decrease with the driving velocity. 
A question appears, if the oscillations are of stick-slip or pure slip kind? The range of 
velocities, which are of order of $\mu m/s$, do not allow to observe the stick effect directly. 
We note that in another experiment with PMMA (\cite{bri}, Fig.5), the amplitude of the 
oscillations increases with the velocity. On the other hand, one could argue that the validity 
of analytical calculations in Ref. \cite{thf} is limited by the assumed shape of the 
function $F(v)$, which was a polynomial function of third order. 

The aim of this paper is to provide new argument in this discussion by numerical calculations, 
for the shape of the function $F(v)$ other than in Ref. \cite{thf}. After 
Ref. \cite{vieira}, we assume this function to be proportional to $(1+av)^{-1}$. Alternatively 
we add a linear term $bv$ to the same function, to reproduce the transition from the stick-slip 
to the uniform motion. However, here we are interested only in the stick-slip phase. The goal is 
to check if the amplitude of the oscillations increases or decreases with the driving velocity 
$v_0$. 

In the next section we present the assumed equations of motion, the model parameters and the 
numerical results. Conclusions which can be derived from  these results are presented in 
Section III. The text is closed by the above remarked Appendix on the Rice-Ruina model.

\section{The model and the results}

For two versions of the formula for the friction force $F_i(v)$, $i=1,2$, the equation of 
motion is

\begin{equation}
m \ddot{x}=-k(x-v_0 t)-F_i(\dot{x})
\end{equation}
where $m$ and $x$ are the block mass and the block position, $k$ is the spring constant, 
$v_0t$ is the position of the driving mechanism, $F_1(v)=\mu_0W/(1+av)$, $F_2=F_1+bv$, 
and $\mu_0W=F_i(0)$ is the static friction force. This equation is equivalent to a dimensionless 
equation

 \begin{equation}
\ddot{u}=\nu \tau -u -\frac{1}{1+cu}-pu
\end{equation}
where $\tau=\sqrt{\frac{k}{m}}t$, $u=\frac{kx}{\mu_0W}$, $\nu=\beta v_0$, 
$\beta=\frac{\sqrt{km}}{\mu_0W}$, $\dot{x}=\frac{dx}{d\tau}$, $c=a/\beta$ and $p=b/\beta$. 
This version, controlled by two $(F=F_1)$ or three $(F=F_2)$ parameters, is used for the numerical 
calculations. The values of the parameters can be reproduced at least partially from the 
experimental data \cite{zv}. The normal load $W=196$ $mN$, the velocity $v_0$ is from $5$ 
to $1000$ $\mu m/s$. From the fitting of the results to the time dependence of the spring 
force (Fig. 2 in Ref. \cite{zv}) we can get $\mu_0=0.46$, $a=160$ $s/m$ and $k=62$ $N/m$. 
The sample mass is about $1$ $g$, but it is certainly much less that the mass $m$ of the 
movable part of the setup, which cannot be deduced from the content of Ref. \cite{zv}. 
It seems reasonable to evaluate its order of magnitude to be about $1000$ $g$. In Fig. 1 
we show the time dependence of the spring force. This is to be compared with Fig. 2 of 
Ref. \cite{zv}. The accordance is good. This plot proves only that the fit is successful.

\begin{figure}
\begin{center}
\includegraphics[angle=0,width=.6\textwidth]{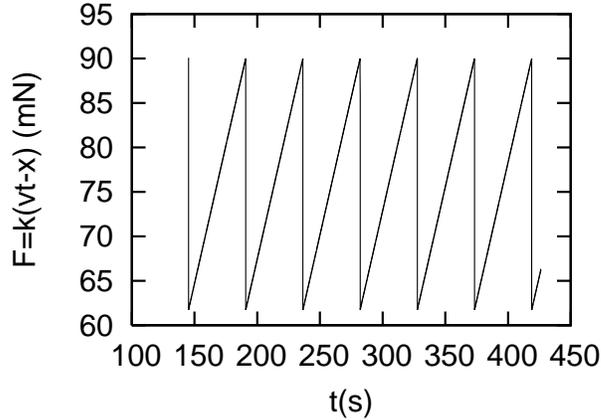}
\caption{ Calculated time dependence of the spring force. 
The range of force is taken as the amplitude. 
The results are to be compared with Fig. 2 in Ref. \cite{zv}.}
\label{Fig.1}
\end{center}
\end{figure}

However, the calculated velocity dependence of the amplitude of the spring force is entirely 
different from the experimental data. Both the  curve obtained for $b=0$ and the experimental 
points are shown in Fig. 2. Other calculated curves are the driving velocity dependence of the 
period of oscillations and the period and amplitude dependences on the normal load. These 
curves agree with the experimental data in the sense that they increase or decrease in the 
same way as in experiment. More detailed comparison is difficult, because we do not know 
the parameters: the mass $m$ and the coefficients $a$ and $b$. However, the results of the 
simulations make clear that the calculated amplitude increases with the driving velocity. 
The only exception is when the mass is very small, less than $1$ $g$, but even in this case 
the obtained dependence is very weak and incomparable with the experimental data.

In the same Fig. 2 we show also the data for $p=2.38$. This value is small enough to keep 
the range of the driving velocity in the stick-slip regime. However, our qualitative 
conclusions remain unchanged.

A comparison of our results with the experimental data of Ref. \cite{bri} can be only qualitative.
The experimental setup is different from a simple block plus spring. However, some 
quantities are analogous to the previous case. For example, the spring constant is substituted 
by an elastic constant of the PMMA sample. On the other hand, numerical value of the coefficient
$\beta$ remains unknown. The value of $a$ is about $250$ $s/m$ \cite{bri}. The coefficient $b$ is 
set to zero. In Fig.3 we show the amplitude dependence on the driving velocity for 
three values of $\beta$: the same as in Ref. \cite{zv} ($\beta=42$ $s/m$),  $158$ $s/m$ and 
$418$ $s/m$. The amplitude is in arbitrary units.  For all investigated cases (three shown plus 
several others) the amplitude either remains constant or increases with $v_0$.
The increase of the experimental results of Ref. \cite{bri} can be reproduced for $\beta=158$ $s/m$.
The vertical scale of each curve depends on the coefficient $\mu _0 W$.

\section{Conclusions}

Summarizing the experimental data \cite{zv,bri}, we are faced with two experiments with 
opposite behaviour of the amplitude of the oscillations with the driving velocity. In the 
case of Ref. \cite{zv}, the amplitude decreases about two times when the velocity increases 
by two orders of magnitude. In the case of Ref. \cite{bri}, it increases about two times 
when the velocity increases by one order of magnitude. This qualitative difference of behaviour 
calls for an explanation with qualitatively different mechanisms.

We postulate a solution that the dependence of the amplitude on the driving velocity, as 
calculated in Ref. \cite{thf}, should be treated as a criterion to distinguish between the 
stick-slip and pure slip oscillations. With this criterion, the data of Ref. \cite{zv} should be 
assigned to pure slip oscillations, and the data of Ref. \cite{bri} - to the stick-slip phase.

{\bf Acknowledgements} The authors are grateful to Tristan Baumberger, Piotr Grabowski and 
Wojciech S{\l}omczy\'nski for helpful discussions and remarks. 

{\bf Appendix}

The equations of motion in the Rice-Ruina model are

\begin{equation}
\frac{k}{W}(v_0t-x)=\mu_0+B\ln \frac{\Phi}{\Phi_0}+A\ln \frac{\dot{x}}{V_r}
\end{equation}

\begin{equation}
\dot{\Phi}=1-\frac{\dot{x}\Phi}{D_0}
\end{equation}
where $x-v_0t$ is the block position with respect to the driving mechanism, $v_0$ is the 
driving velocity, $k$ is the spring constant, $W$ is the normal force, $\mu _0$ is a reference 
value of the friction coefficient for steady sliding at some velocity $V_r$. During sliding, 
the microcontacts are refreshed, on average, after a distance $D_0$. The state of these 
microcontacts is described by the variable $\Phi $, which interpolates between the time of stick for the 
block sticked and $D_0/v_0$ for steady sliding. Finally, $\Phi_0=D_0/V_r$ and $A$, $B$ are 
unitless material constants. We note that $B>A$ in the experimental data \cite{bau2}.

The equations can be transformed to an autonomous form. Denoting $(x-v_0t)/D_0=\alpha$, 
$\Phi /\Phi _0=\phi$, $V_rt/D_0=\tau$, $V/V_r=\omega$, exp$(-\mu _0/A)=\gamma$, 
$KD_0/W=\kappa$, we get dimensionless equations

\begin{equation}
\dot{\alpha}=-\omega+\gamma\phi^{-\frac{B}{A}}\exp [-\frac{\kappa}{A}\alpha]
\end{equation}

\begin{equation}
\dot{\phi}=1-\gamma\phi^{1-\frac{B}{A}}\exp [-\frac{\kappa}{A}\alpha]
\end{equation}
where the time derivative is over $\tau$. To apply the Dulac criterion \cite{glen}, we need 
an auxiliary function $g(\alpha,\phi)$; here the simplest $g=1$ is sufficient. The divergence is 

\begin{equation}
(B-A-\kappa)\frac{\gamma}{A}\phi^{-\frac{B}{A}}  \exp [-\frac{\kappa}{A}\alpha]
\end{equation}
and it is different from zero except at the transition point from the uniform to the stick-slip 
motion. Then, there is no periodic orbits if $B-A-\kappa \ne 0$.

\newpage

\begin{figure}
\begin{center}
\includegraphics[angle=0,width=.6\textwidth]{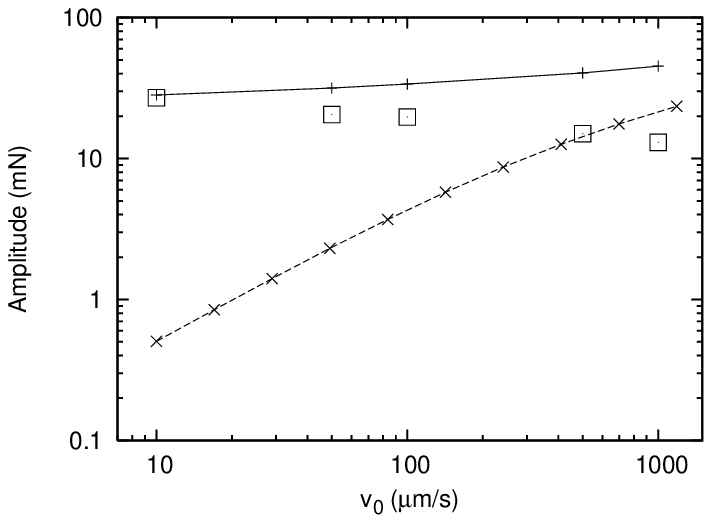}
\caption{Amplitude of the friction force as dependent on the driving velocity $v_0$.
Calculations are performed for $b=0$ (solid line) and $b=100$ $s/m$ (dotted line). 
Experimental points are taken from Fig. 4 in Ref. \cite{zv}.}
\label{Fig.2}
\end{center}
\end{figure}

\begin{figure}
\begin{center}
\includegraphics[angle=0,width=.6\textwidth]{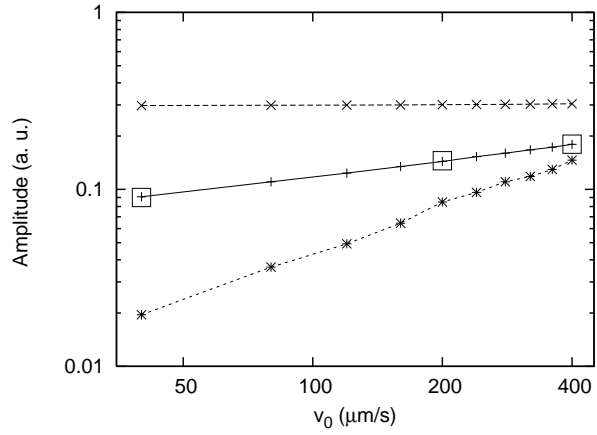}
\caption{Amplitude of the friction force as dependent on the driving velocity $v_0$.
Calculations are performed for $b=0$  and three different values of $\beta$, from 
$42$ $s/m$ (highest curve) till $418$ $s/m$ (lowest curve). Experimental points 
are taken from Fig. 4 in Ref. \cite{bri}.}
\label{Fig.3}
\end{center}
\end{figure}

\end{document}